\documentstyle[prb,aps,epsf,floats]{revtex}
\begin{document}
\renewcommand{\textfraction}{0.0}
\renewcommand{\floatpagefraction}{.7}
\setcounter{topnumber}{5}
\renewcommand{\topfraction}{1.0}
\setcounter{bottomnumber}{5}
\renewcommand{\bottomfraction}{1.0}
\setcounter{totalnumber}{5}
\setcounter{dbltopnumber}{2}
\renewcommand{\dbltopfraction}{0.9}
\renewcommand{\dblfloatpagefraction}{.7}

\draft

\title{Fixed Points of Hopfield Type Neural Networks}

\author{Leonid B. Litinskii }
\address{High Pressure Physics Institute of Russian Academy of Sciences
Russia, 142092 Troitsk Moscow region, e-mail: litin@hppi.troitsk.ru}

\date{May 3, 1999}
\maketitle

\begin{abstract}
The set of the fixed points of the Hopfield type network is under
investigation. The connection matrix of the network is constructed according to
the Hebb rule from the set of memorized patterns which are treated as 
distorted copies of the standard-vector. It is found that the dependence of 
the set of the fixed points on the value of the distortion parameter can be 
described analytically. The obtained results are interpreted in the terms of 
neural networks and the Ising model.
\end{abstract}
\section{INTRODUCTION}
The problem of maximization of a symmetric form 
$$\left\{\begin{array}{l} F(\vec\sigma)=\sum_{i,j=1}^n J_{ij}\sigma_i\sigma_j
\to {\rm max},\ \sigma_i=\{\pm 1\},\\
\vec\sigma=(\sigma_1,\ldots,\sigma_n),\ J_{ij}=J_{ji},\ i,j=1,2,\ldots,n,
\end{array}\right.\eqno(1)$$
which is quadratic in {\it spin} variables $\sigma_i$, is under investigation.
This problem arises in the Ising model, in the  
surface physics, in the theory of optimal coding, in factor analysis, in the 
theory of neural networks and in the optimization problems 
 \cite{Hopf,Dom,Surf1,Sour,Par,Brav2,Lit1}. The aim is to obtain an  
effective method for the search of the global maximum of the functional and 
a constructive description of the set of its local maxima.

The solution of the problem (1) is sought among the $n$-dimensional vectors 
$\vec\sigma$. These vectors define $2^n$ configurations; in what follows they
are called {\it the configuration vectors}. The configuration vector
which gives the solution of the problem (1) is called {\it the ground state}.

Let's settle with the choice of the diagonal elements of the matrix
${\bf J}=(J_{ij})_{i,j=1}^n$. Since for any configuration vector the diagonal
gives the same contribution to the value of the functional, the local maxima 
of
the functional do not depend on $J_{ii}$. Consequently, $J_{ii}$ can be chosen
in an arbitrary suitable form. In the same time, the set of the local maxima 
of
the functional $F$ is the same as the set of the fixed points of a neural
network whose connection matrix differs from $\bf J$ only by zero diagonal 
elements\footnote{For details see \cite{Lit3}.}. In what follows, we call
$\bf J$ {\it the connection  matrix} irrespective of its diagonal. And with
regard to neural networks we have in mind that its diagonal elements are equal 
zero.

Without the loss of generality $\bf J$ can be considered as a positive
semidefinite matrix. (It always can be achieved by a suitable choice of its
diagonal elements.) Then the matrix $\bf J$ can be presented as a product
 $${\bf J} ={\bf S}^{\rm T}\cdot\bf S,\eqno(2)$$
where $\bf S$ is a $(p\times n)$-matrix with {\it real} matrix elements. The
number of its rows is equal to the rank of the matrix $\bf J$ and it does not
exceed $n-1$:
$$p=\mbox{rank }{\bf J}\le n-1.$$
The representation (2) for the connection matrix plays an important role in 
the
theory of neural networks. If all the elements of the matrix $\bf S$ are equal
$\{\pm 1\}$, the connection matrix is called the Hebb matrix and the related
neural network is called the Hopfield network. Of course, in general case
when according to Eq. (2) the matrix $\bf S$ is obtained from an {\it
arbitrary}  
symmetric matrix $\bf J$, not necessarily all the elements of $\bf S$ 
are equal $\{\pm 1\}$. By analogy with the Hopfield model, in this case Eq.(2)
is called {\it the Hebb type} representation and the related neural network is
called {\it the Hopfield type} network. Summarizing, we can say that the local
maxima of the problem (1) are the fixed points of a Hopfield type network.

We investigate the problem (1) in the case of the
connection matrix constructed with regard to the Hebb-like rule (2) from 
the $(p\times n)$-matrix $\bf S$  of the form
$${\bf S}=\left(\begin{array}{ccccccc}
1-x&1&\ldots&1&1&\ldots&1\\
1&1-x&\ldots&1&1&\ldots&1\\
\vdots&\vdots&\ddots&\vdots&\vdots&\ldots&\vdots\\
1&1&\ldots&1-x&1&\ldots&1\end{array}\right), \eqno(3.a)$$
where $x$ is an arbitrary real number. The connection matrix $\bf J$ has the 
form:
$${\bf J}={\bf S}^{\rm T}{\bf\cdot S}=\left(\begin{array}{ccccccc}
p-2x+x^2&p-2x&\ldots&p-2x&p-x&\ldots&p-x\\
p-2x&p-2x+x^2&\ldots&p-2x&p-x&\ldots&p-x\\
\vdots&\vdots&\ddots&\vdots&\vdots&\vdots&\vdots\\
p-2x&p-2x&\ldots&p-2x+x^2&p-x&\ldots&p-x\\
p-x&p-x&\ldots&p-x&p&\ldots&p\\
\vdots&\vdots&\vdots&\vdots&\vdots&\ddots&\vdots\\
p-x&p-x&\ldots&p-x&p&\ldots&p\end{array}\right).\eqno (3.b) $$
According to the conventional neural network tradition, we treat the
$n$-dimensional vectors $\vec s^{(l)}$, which are the rows of the matrix 
$\bf S$, as $p$ {\it generalized memorized patterns} embedded in the network
memory. Then the following meaningful interpretation of the problem can be 
suggested: 
{\it the network had to be learned by p-time showing of the standard 
$$\vec\varepsilon (n)= (1,1,\ldots,1),$$ 
but an error crept into the learning process and in fact the network was 
learned with the help of $p$ distorted copies $\vec s^{(l)}$ of the standard 
$\vec\varepsilon (n)$:
$$\vec s^{(l)}=(1,\ldots,1,\underbrace{1-x}_l,1,\ldots,1),\quad l=1,2,\ldots,
p.\eqno(4)$$} 
The problem (1)-(3) will be called the {\it basic model}. In what follows we
present its generalization.

When $x$ is equal zero, the network is found to be learned by $p$ copies of 
the standard $\vec\varepsilon (n)$. It is well-known that in this case the 
vector $\vec\varepsilon (n)$ itself is the ground state and the functional has
no other local maxima. For continuity reasons, it is clear that the same is 
true for a sufficiently small distortion $x$. But when $x$ increases the 
ground
state changes. For the problem (1)-(3) we succeeded in obtaining the 
analytical
description of the dependence of the ground state on the value of the 
distortion parameter. 

In \ref{2} the main mathematical results are presented; in \ref{3} we give
the interpretation of the obtained results in terms of the neural network 
and the Ising model. Some proofs which are not significant for the
understanding of the structure of the solution enter Appendix.

We would like to mention that our approach is very close to the problem of 
generalization in neural networks \cite{Fon,Kre,Sil}. Under the classical
setting of the problem, one is investigating the network ability to 
reconstruct the set of memorized patterns if during the learning their
distorted copies have been used. In contrast to the classical setting, in our
case the distortions are deterministic ones. Moreover, the value of the 
distortion $x$ is the same for all the memorized patterns and only one 
coordinate is distorted every time. In \ref{2} it is shown that some of 
these
restrictions can be omitted. However, the most important restriction of our
approach is investigation of the generalization in the case
of one memorized pattern only, but not the 
set of memorized patterns. Nevertheless, the meaningful interpretation of the
obtained results (see \ref{3}) is of interest for the more general case.

{\bf Notations.} 
We denote by $\vec \varepsilon (k)$ the configuration vector 
which is collinear to the bisectrix of the principle orthant of the space 
$\rm R^k$. The vector, which after $p$ distortions 
generates the set of the memorized patterns $\vec s^{(l)}$, is called 
{\it the standard-vector} or {\it the standard}. Next, $n$ is the number of 
the spin variables, $p$  
is the number of the memorized patterns and $q=n-p$ is the number of the 
nondistorted coordinates of the standard. Configuration vectors are 
denoted by small Greek letters. We use small Latin letters to denote vectors 
whose coordinates are real. The n-dimensional vectors are numerated by 
superscripts enclosed in brackets and their coordinates are numerated by 
subscripts. Matrices are denoted by half-bold capital Latin letters.

\section{Mathematical Results}
\label{2}
Let us look for the local maxima among configuration  vectors 
whose last coordinate is positive.

\subsection{Basic Model}
\label{2.1}
Since $q$ last columns of the matrix $\bf S$ are the same, the configuration
vector which is "under the suspicion" to provide an extremum is of the 
form\footnote{For details see \cite{Lit2}.}
$$\vec\sigma^*=(\underbrace{\sigma_1,\sigma_2,\ldots,\sigma_p}_
{\vec\sigma^\prime},\underbrace{1,\ldots,1}_q),\eqno(5)$$
where {\it we denote by $\vec\sigma^\prime$ the $p$-dimensional part of the 
vector $\vec{\sigma}^{*}$, which is formed by its first $p$ coordinates.}
The direct calculations (or see \cite{Lit5}) show that  
$$F(\vec \sigma^*)\propto x^2 - 2x(q+p\cos w)\cos w +(q+p\cos w)^2,\eqno(6)$$
where
$$\cos w=\frac{\sum_{i=1}^p \sigma_i}p=\frac{(\vec\sigma^\prime,
\vec\varepsilon (p))}{\parallel\vec\sigma^\prime\parallel\cdot\parallel
\vec\varepsilon (p)\parallel}\eqno(7)$$
is the cosine of the angle between $p$-dimensional vectors $\vec\sigma^\prime$ 
and $\vec\varepsilon (p)$. Depending on the number of the negative coordinates
of the  vector $\vec\sigma^\prime$, $\cos w$ takes the values
$$\cos w_k=1-\frac{2k}{p},\quad k=0,1,\ldots,p.$$
Consequently, $2^p$ "suspicious-looking" vectors $\vec\sigma^*$ are 
grouped into the $p+1$ {\it classes } $\Sigma_k$ and
the functional $F(\vec\sigma^*)$ has the same value $F_k$ for all the 
vectors from the same class:
$$F(\vec\sigma^*)\equiv F_k\quad \forall \vec\sigma^*\in\Sigma_k.$$ 
The classes $\Sigma_k$ are numerated by the number $k$ of the negative 
coordinates which have the relevant vectors $\vec\sigma^*$. The 
number of the vectors in the class $\Sigma_k$ is equal to $(^p_k)$.

To find the ground state under a given value of $x$, it is necessary to
determine the greatest of the values $F_0(x),F_1(x),\ldots,F_p(x)$. Under the
comparison the common term $x^2$ can be omitted. Therefore, to find out how 
the ground state depends on the parameter $x$, it is necessary to examine the 
family of the straight lines
$$L_k(x)=(q+p\cos w_k)^2 - 2x(q+p\cos w_k)\cos w_k. \eqno(8)$$
In the region where the $L_k(x)$ majorizes all the other straight lines, the 
ground state belongs to the class $\Sigma_k$ and is $(^p_k)$-times degenerate.
The analysis of the relative position of the straight lines $L_k(x)$ is given
in Appendix and leads to the following 

\noindent
{\bf Theorem.} {\it As $x$ varies from $-\infty$ to $\infty$ the ground state 
in consecutive order belongs to the classes 
$$\Sigma_0,\Sigma_1,\ldots,\Sigma_{k_{max}}.$$
The rebuilding of the ground state from the class $\Sigma_{k-1}$ into the class
$\Sigma_k$ occurs at the point $x_k$ of intersection of the straight lines
$L_{k-1}(x)$ and $L_k(x)$:
$$x_k=\frac{p}{2}\left(1+\frac{q}{q+p(\cos w_{k-1}+\cos w_k)}\right)=
p\cdot\frac{n-(2k-1)}{n+p-2(2k-1)},\quad k=1,2,\ldots,k_{max}.\eqno(9)$$
If $\frac{p-1}{n-1}<\frac13$, one after another all the $p$ rebuildings 
of the ground state take place according to the
above scheme: $k_{max}=p$. And if $\frac{p-1}{n-1}>\frac13 $, the last
rebuilding is the one with the number $k_{max}=\left[\frac{n+p+2}4\right]$.
The functional has no other local maxima.}

The Theorem allows to solve a lot of practical problems. For example, let $n$ 
and $p$ be fixed. Then, for any preassigned $x$ it is sufficient to know 
between which $x_k$ and $x_{k+1}$ it finds itself to determine which of the 
classes $\Sigma_k$ provide the global maximum of the functional 
$F(\vec\sigma)$, what is its value and the degeneration of the ground state. 
On
the contrary, for the fixed $x$, it is possible to find out such $n$ and $p$ 
for which the ground state belongs to the preassigned class $\Sigma_k$, 
{\it etc.}

\begin{figure}[htb]
\begin{center}
\leavevmode
\epsfxsize = 16.2truecm
\epsfysize = 7.5truecm
\epsffile{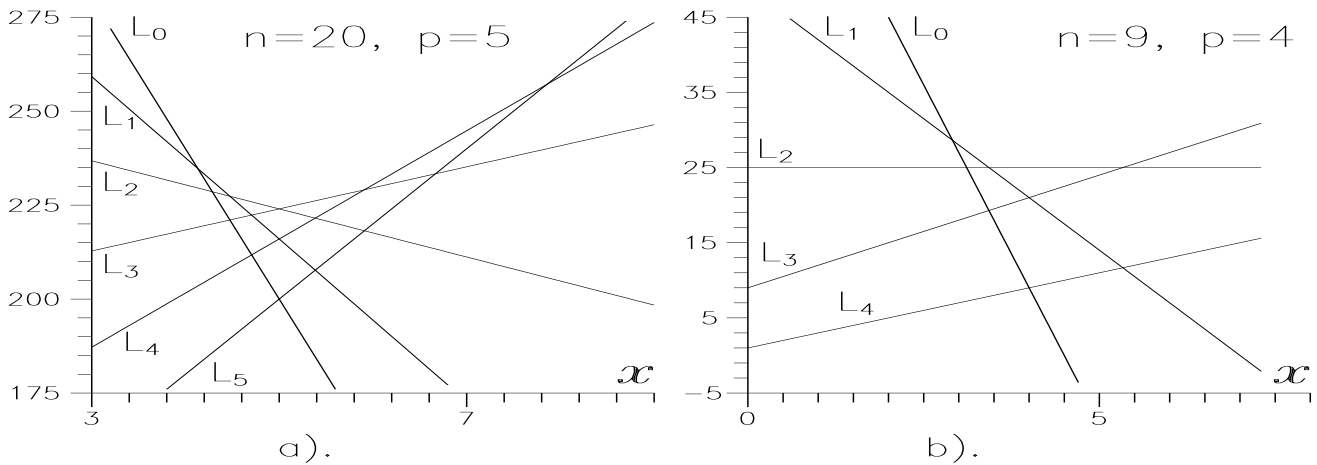}
\caption{The typical behavior of the straight lines 
$L_k(x),\ k=0,1,\ldots,p$. The rebuildings of the ground state occurs at the
points $x_k$ of the intersection of the straight lines $L_{k-1}$ and $L_k$.
Inside the interval $(x_k,x_{k+1})$ the ground state belongs to the class
$\Sigma_k$. When $x$ increases: a). all the rebuildings of the ground state
occur 
($k_{max}=5$), because $\frac{p-1}{n-1}<\frac13$; b). only 
$k_{max}=\left[\frac{n+p+2}4\right]=3$ rebuildings of 
the ground state occur, because
$\frac{p-1}{n-1}>\frac13$.}
\end{center}
\end{figure}

The points $x_k$ (9), where the rebuildings of the ground state take place will
be called the {\it rebuilding points}.

In Fig.1 the typical examples of the relative position of the straight lines 
$L_k(x)$ are presented for the cases $\frac{p-1}{n-1}<\frac13$ ({\it a}) 
and $\frac{p-1}{n-1}>\frac13$ ({\it b}). When $x$ changes from $-\infty$ to 
$x_1$, the ground state is the standard-vector $\vec\varepsilon (n)$ (it 
exhausts the class $\Sigma_0$). In the point $x_1$ the ground state rebuilds
from 
the class $\Sigma_0$ to the class $\Sigma_1$ and becomes $p$-times degenerate.
When $x$ achieves the value $x_2$, the ground state rebuilds from the class 
$\Sigma_1$ to the class $\Sigma_2$ and becomes $p\choose 2$-times degenerate, 
and so on. As $x$ increases, at first the value of the functional 
{\it for the ground state} monotonically decreases and then, after reaching 
the minimum value, increases monotonically. For what follows let us note that 
if $p<n-1$, then $k_{max}\ge [\frac{p+1}2]$ and $x_{k_{max}}\ge p$.

The case $p=n$ worth to be specially mentioned. Here all the rebuilding points 
$x_k$ stick to one point
$$x^\prime\equiv x_k =\frac{n}2,\quad k=1,2,3,\ldots,\left[\frac{n+1}2\right].
\eqno(10) $$
For any $x$ from the left of $x^\prime$ the ground state is the 
standard-vector
$\vec\varepsilon (n)$, and for $x$ from the right of $x^\prime$ the ground 
state belongs to the class $\Sigma_{[\frac{n+1}2]}$ and is 
$n\choose [\frac{n+1}2]$-times degenerate. 

The interval $x_1 < x < x_{k_{max}}$ will be called {\it 
the rebuilding region} of the ground state. This region is examined in details
in Appendix. Here we would like to mention only, that its left boundary 
$x_1\ge\frac{p}2$ and it is the monotonically increasing function of $p$ as
well as of
$n$. And also, when $p=const$ and $n\to\infty$ the rebuilding region contracts 
to the point 
$$x^{\prime\prime}=p.\eqno(11)$$
In this case for  $x<x^{\prime\prime}$ the ground state is the standard-vector
and for $x>x^{\prime\prime}$ the ground state belongs to the class $\Sigma_p$; 
again it is a nondegenerate one.  

\subsection{Generalization of the Basic Model} 
\label{2.2}
Let's examine some generalizations of the basic model, which seem to be the
most important. From our point of view they significantly widen the class of
possible connection matrices for which the 
effective solution of the problem (1) can be obtained.

\subsubsection{An arbitrary cofiguration standard-vector}
\label{2.2.1}
An arbitrary configuration vector 
$$\vec{\alpha}=(\alpha_1,\ldots,\alpha_p,\alpha_{p+1},\ldots,\alpha_n)$$ 
can be used in place of the standard $\vec\varepsilon (n)$, and its $p$
distorted copies of the type (4) can be used as memorized patterns. These
changes affect the elements of the connection matrix:
$$J_{ij}^{(\alpha)}=J_{ij}\alpha_i\alpha_j,\ i,j=1,2,\ldots,n.\eqno(12)$$
If we introduce a diagonal $(n\times n)$-matrix
$${\bf A}=\mbox{diag 
}(\alpha_1,\ldots,\alpha_p,\alpha_{p+1}\ldots,\alpha_n),$$
then ${\bf J}^{(\alpha)}={\bf A}\cdot{\bf J}\cdot{\bf A}$ and the functional 
we
have to maximize, takes the form:
$$F^{(\alpha)}=({\bf J^{(\alpha)}}\vec\sigma,\vec\sigma)=({\bf J}\cdot
{\bf A}\vec\sigma,{\bf A}\vec\sigma)=({\bf J}\vec\xi,\vec\xi)=F(\vec\xi).$$ 
Here $\vec\xi={\bf A}\vec\sigma$, and the functional $F(\vec\xi)$ has been
examined above. Consequently, all the results of the basic model remain 
correct, though here they are relative to configuration vectors
$$\vec\sigma^*=(\alpha_1\sigma_1,\alpha_2\sigma_2,\ldots,\alpha_p\sigma_p,
\alpha_{p+1},\ldots,\alpha_n).$$
Thus, $\vec\varepsilon (n)$ can be used as the standard-vector without the 
loss
of generality. 

\subsubsection{Rotation of the memorized patterns}
\label{2.2.2}
An interesting generalization of the basic model is obtained when the 
memorized patterns (4) are subjected to rotation as a whole. Let the 
nontrivial
part of the rotation matrix ${\bf U}=(u_{ij})$ has the dimension $p\times p$.
It acts on the first $p$ coordinates of $n$-dimensional vectors 
transforming the distorted part of the standard into a $p$-dimensional 
vector $\vec u$:
$$\vec u={\bf U}\cdot\vec\varepsilon (p)=\left(\begin{array}{c}u_1\\u_2\\
\vdots\\u_p\end{array}\right),\quad
u_l=\sum_{i=1}^p u_{li},\mbox{ }l=1,2,\ldots,p,\mbox{ }\parallel\vec 
u\parallel^2=p.\eqno(13)$$
As a result of the rotation the $l$-th memorized pattern takes the form
$$(u_1-xu_{1l},u_2-xu_{2l},\ldots,u_p-xu_{pl},1,\ldots,1),\eqno(14)$$
and it is convenient to choose the elements of the relevant connection matrix
${\bf J}^{(U)}$ in the form  
$$J_{ij}^{(U)}=J_{ij}u_iu_j,\ i,j=1,2,\ldots,n.\eqno(15)$$
From Eqs. (14), (15) it directly follows that if the standard-vector 
$\vec\varepsilon (n)$ is not changed as a result of the rotation (in other 
words, if $u_1=u_2=\ldots=u_p=1$) then the connection matrix ${\bf J}^{(U)}$ 
coincides with the matrix $\bf J$. Consequently, all the results of the 
basic model are valid, though in this case the memorized patterns (14) are 
obtained from $\vec\varepsilon (n)$ by simultaneous (!) distortion of its $p$ 
coordinates.

But if the rotation results in $u_l$ which are not all equal unity, the 
standard-vector shifts. Then it is convenient to introduce a diagonal
$(n\times n)$-matrix 
$${\bf D}=(u_1,u_2,\ldots,u_p,1,\ldots,1)$$
and present the connection matrix ${\bf J}^{(U)}$  as a triple product 
${\bf J}^{(U)}={\bf D}\cdot{\bf J}\cdot{\bf D}$. By analogy with the above 
reasonings it is easy to show that again only the configuration vectors 
$\vec\sigma^*$ (5) are "under the suspicion" to provide an extremum of the 
functional $F^{(U)}(\vec\sigma)=({\bf J}^{(U)}\vec\sigma,\vec\sigma)$. Again 
the value of the functional $F^{(U)}(\vec \sigma^*)$ is given by Eq.(6), but 
$\cos w$ has to be calculated with the help of other equation:
$$\cos  w=\frac{\sum_{i=1}^p \sigma_i\cdot u_i}p=
\frac{(\vec\sigma^\prime,\vec u)}{\parallel \vec \sigma^\prime 
\parallel \cdot \parallel \vec u \parallel}\eqno(16)$$
(compare with Eq.(7)). Thus, the value of the functional 
$F^{(U)}(\vec\sigma^*)$ is completely defined by the distortion $x$ and the
value of the cosine of the angle between $\vec\sigma^\prime$ and $\vec u$. 
Consequently, again the vectors $\vec\sigma^*$ are grouped in the classes 
$\Sigma^{(U)}_k$, inside which the value of the functional 
$F^{(U)}(\vec\sigma^*)$ is constant. The vectors $\vec\sigma^*$ belong to the 
same class if their $p$-dimensional parts are equidistant from the vector 
$\vec u$. The number of the classes $\Sigma^{(U)}_k$ is equal to the number of
{\it the different} values of $\cos w$ (16).

Thus, it is necessary to put in order $2^p$ vectors $\vec\sigma^\prime$ with 
regard to their proximity to the vector $\vec u$. This problem will be 
discussed in another publication. But now let us assume that the desired 
ordering has been done and a decreasing sequence of the values of $\cos w$ 
(16) has been obtained: 
$$\cos w_0>\cos w_1>\ldots>\cos w_t.\eqno(17)$$
Here the number of the classes $\Sigma^{(U)}_k$ is equal to $t+1$. We 
denote by $v_k$ the number of the vectors $\vec\sigma^*$ in the  $k$-th class:
$$v_0+v_1+\ldots+v_t=2^p.$$ 
Since the half of the vectors $\vec\sigma^\prime$ can be
obtained from the other half of the full set of these vectors by changing
signs of all their coordinates, for every $k$ we have obvious equalities:
$$\cos w_k=-\cos w_{t-k},\quad v_k=v_{t-k}.$$ 
Therefore, $\cos w_k$ is negative beginning from some number $k$. 

We see that as above to determine the dependence of the ground 
state on the parameter $x$, it is necessary to examine the family of 
the straight lines $L_k(x)$ (8) where the cosines are calculated according to
Eq.(16). The analysis of the placement of the straight 
lines gives the result which generalizes the Theorem (the proof is given in
Appendix):  

\noindent
{\it As $x$ varies from $-\infty$ to $\infty$ the ground state 
in consecutive order belongs to the classes 
$$\Sigma^{(U)}_0,\Sigma^{(U)}_1,\ldots,\Sigma^{(U)}_{k_{max}}.$$
The rebuilding of the ground state from the class $\Sigma^{(U)}_{k-1}$ into the
class
$\Sigma^{(U)}_k$ occurs at the point $x_k$ of intersection of the straight 
lines $L_{k-1}(x)$ and $L_k(x)$:
$$x_k=\frac{p}{2}\left(1+\frac{q}{q+p(\cos w_{k-1}+\cos w_k)}\right),
\quad k=1,2,\ldots,k_{max},\eqno(18)$$ 
where $\cos w_k$ belong to the ordered sequence
(17). If $x_1>\frac34p$, one after another all the $t$ rebuildings 
of the ground state take place according to the
above scheme: $k_{max}=t$. And if $x_1<\frac34p$, the rebuildings of the 
ground
state come to an end when the denominator in Eq.(18) becomes negative.}

{\bf Note 1.} When the distortion $x$ belongs to the interval $(x_k,x_{k+1})$,
all the configuration vectors from the class $\Sigma^{(U)}_k$ are 
the local maxima of the functional $F^{(U)}(x)$. At the same time the set of
these vectors is the set of the fixed points of the Hopfield type network 
whose connection matrix is ${\bf J}^{(U)}$ (we remind that its diagonal
elements are equal zero). But the composition of each of the classes 
$\Sigma^{(U)}_k$ is 
determined by the values of $\{u_l\}_{l=1}^p$ (13) only. And the choice  of
$\{u_l\}_{l=1}^p$ is completely in the researcher's hand. In other words,
selecting 
$\{u_l\}_{l=1}^p$ and the distortion parameter $x$, the Hopfield type network 
with a preassigned set of fixed points can be created. 

Let's explain this statement with the help of a simple example. Let's create
the network with the only fixed points
$$\begin{array}{rrrrrrrrrrl}\vec\alpha=(&-1,&-1,&-1,&1,&1,&1,&1,&\ldots,&1&)\\
\vec\beta=(&1,&1,&1,&-1,&-1,&1,&1,&\ldots,&1&)\\
\vec\gamma=(&1,&1,&1,&1,&1,&-1,&1,&\ldots,&1&).\end{array}$$
Then it is necessary, firstly, to group these vectors into one class 
$\Sigma^{(U)}_k$, and, secondly, to choose such distortion $x$ which provides
the belonging of the ground state to this class. The first step can be easily
done, if with the help of a rotation of the memorized patterns the standard 
$\vec\varepsilon (n)$ transforms into the vector
$$\vec u=(u_1,u_1,u_1,\frac{3u_1}2,\frac{3u_1}2,3u_1,u_7,\ldots,u_p),\quad  
0<u_1<\frac{3u_1}2<3u_1<u_7<\ldots<u_p.$$

Indeed, it is evident that in this case the first six values of the cosines
(16) and the relevant classes $\Sigma_k^{(U)}$ are:
$$\begin{array}{lll}\cos w_0=\frac{9u_1+\sum_{i=7}^p u_i}p,&
\Sigma_0^{(U)}=&\vec\varepsilon (n)=\begin{array}{rrrrrrrrrrl}(&1,&1,&1,&1,&1,
&1,&1,&\ldots,&1&)\end{array}\\
\cos w_1=\cos w_0-2\cdot\frac{u_1}p,&
\Sigma_1^{(U)}=&\left\{\begin{array}{rrrrrrrrrrl}(&-1,&1,&1,&1,&1,&1,&1,
&\ldots,&1&)\\
(&1,&-1,&1,&1,&1,&1,&1,&\ldots,&1&)\\
(&1,&1,&-1,&1,&1,&1,&1,&\ldots,&1&)\end{array}\right.\\
\cos w_2=\cos w_0-3\cdot\frac{u_1}p,&
\Sigma_2^{(U)}=&\left\{\begin{array}{rrrrrrrrrrl}(&1,&1,&1,&-1,&1,&1,&1,
&\ldots,&1&)\\
(&1,&1,&1,&1,&-1,&1,&1,&\ldots,&1&)\end{array}\right.\\
\cos w_3=\cos w_0-4\cdot\frac{u_1}p,&
\Sigma_3^{(U)}=&\left\{\begin{array}{rrrrrrrrrrl}(&-1,&-1,&1,&1,&1,&1,&1,&
\ldots,&1&)\\
(&1,&-1,&-1,&1,&1,&1,&1,&\ldots,&1&)\\
(&-1,&1,&-1,&1,&1,&1,&1,&\ldots,&1&)\end{array}\right.\\
\cos w_4=\cos w_0-5\cdot\frac{u_1}p,&
\Sigma_4^{(U)}=&\left\{\begin{array}{rrrrrrrrrrl}(&-1,&1,&1,&-1,&1,&1,&1,&
\ldots,&1&)\\
(&-1,&1,&1,&1,&-1,&1,&1,&\ldots,&1&)\\
(&1,&-1,&1,&-1,&1,&1,&1,&\ldots,&1&)\\
(&1,&-1,&1,&1,&-1,&1,&1,&\ldots,&1&)\\
(&1,&1,&-1,&-1,&1,&1,&1,&\ldots,&1&)\\
(&1,&1,&-1,&1,&-1,&1,&1,&\ldots,&1&)\\
\end{array}\right.\\
\cos w_5=\cos w_0-6\cdot\frac{u_1}p,&
\Sigma_5^{(U)}=&\left\{\begin{array}{rrrrrrrrrrl}(&-1,&-1,&-1,&1,&1,&1,&1,&
\ldots,&1&)
\\
(&1,&1,&1,&-1,&-1,&1,&1,&\ldots,&1&)\\
(&1,&1,&1,&1,&1,&-1,&1,&\ldots,&1&)\end{array}\right.
\end{array}$$
Thus, the class $\Sigma_5^{(U)}$ consists of the given configuration
vectors $\vec\alpha, \vec\beta\mbox{ and }\vec\gamma$ exactly. Cosequently, if
the distortion parameter $x$ is chosen inside the interval $(x_5,x_6)$, where
$x_k$ 
are defined by Eq.(18), the relevant network of the Hopfield type has the
preassigned set of the fixed points.

We would like to mention that an additional analysis is required to find out 
the limits of our method for the construction of networks with a preassigned
set of the fixed points.

\subsubsection{Different distortions of the memorized patterns}
\label{2.2.3}
The distortions $x_l$ can differ for every of the memorized patterns (4):
$$\vec s^{(l)}=(1,\ldots,1,\underbrace{1-x_l}_l,1,\ldots,1),\quad 
l=1,2,\ldots,p.$$
With the help of the reasoning the same as in \ref{2.1}, it is easy to
show that as above only the configuration
vectors $\vec\sigma^*$ (5) are important. And the value of the functional
$F(\vec\sigma^*)$ is
$$F(\vec\sigma^*)=\sum_{l=1}^px_l^2 - 2(q+p\cos w)
\sum_{l=1}^p\sigma_lx_l+p(q+p\cos w)^2,$$
where again $\cos w$ is given by Eq.(7). Under the comparison we have to 
omit
the term $\sum_{l=1}^px_l^2$. Now we obtain that to determine the ground
state dependence 
on {\it the set} of the distortions $\{x_l\}_1^p$, it is necessary to examine
the family of the linear functions
$$L(\vec\sigma^*,\{x_l\}_1^p)=(q+p\cos w)^2-\frac{2(q+p\cos w)}p
\sum_{l=1}^p\sigma_lx_l.\eqno(19)$$
For this analysis one can try to use a probabilistic approach, considering
$x_l$ as  
independent realizations of a stochastic variable, which, let us say,
is Gaussianly distributed. This way brings us nearer to the classical problem
of the network ability to generalization (see Introduction). We have not
investigate this approach.

On the other hand, we can remain in the framework of the geometric
approach, which is rather fruitful for this problem. Let's show this.
Without loss of generality we assume that the distortions $x_l$ are arranged
in increasing order. And for simplicity we assume them to be positive. Then, 
$$0<x_1<x_2<\ldots<x_p.\eqno(20)$$
As in \ref{2.1} we have
$$\cos w_k=1-\frac{2k}p\quad\Rightarrow\quad q+p\cos w_k=n-2k,\mbox{ }k=0,1,
\ldots,p,$$ 
and the family of the functions (19) decomposes into $p+1$ subfamilies
$L_k(\vec\sigma^*,\{x_l\}_1^p)$. Here the subscript $k$ indicates that the
relevant configuration vector $\vec\sigma^*$ belongs to the class $\Sigma_k$:
$$L_k(\vec\sigma^*,\{x_l\}_1^p)=(n-2k)^2-\frac{2(n-2k)}p
\sum_{l=1}^p\sigma_lx_l,
\quad\vec\sigma^*\in\Sigma_k,\quad k=0,1,\ldots,p.\eqno(21)$$

The decomposition of the family of the functions (19) into $p+1$ subfamilies
(21) allows to find easily the ground state for a given set of the distortions
$\{x_l\}_1^p$. Indeed, to determine the ground state we have, firstly, to find
the vector $\vec\sigma^*(k)$ inside every class $\Sigma_k$, which provides
maximum $L_k^*$ of the function $L_k(\vec\sigma^*,\{x_l\}_1^p)$:
$$L^*_k=L_k(\vec\sigma^*(k),\{x_l\}_1^p)\ge L_k(\vec\sigma^*,\{x_l\}_1^p)
\quad\forall\vec\sigma^*\in\Sigma_k,\quad k=0,1,\ldots,p.$$
And, secondly, we have to determine the greatest among $p+1$ values $L_k^*$. 
The last can be done easily, and the determination of the
vectors $\vec\sigma^*(k)$ does not require additional calculations at all:
if $n-2k>0$, we have to minimize the factor $\sum_{l=1}^p\sigma_lx_l$ (see
Eq.(21)); if $n-2k<0$ the same factor has to be maximized. 
From inequalities (20) it is evident that the factor is minimal at the vector
$$\vec\sigma^*(k)=(\overbrace{1,1\ldots,1,\underbrace{-1,-1,\ldots,-1}_{k}}^
{p},1,1,\ldots,1),$$
and it is maximal at the vector
$$\vec\sigma^*(k)=(\overbrace{\underbrace{-1,-1\ldots,-1}_{k},1,1,\ldots,1}^{p
},1,1,\ldots,1).$$
Thus, when the distortions $x_l$ are different, the geometric approach allows
to determine the ground state very quickly  and efficiently. It seems, that 
the potentiality of the geometric approach  does not exhausted by the mentioned
examples.

{\bf Note 2.} We denote by $\bar x$ the average over all the distortions 
$$\bar x = \frac{\sum_{l=1}^p x_l}p.$$
We write the distortions $x_l$ in the form:
$$x_l=\bar x+d_l,\quad l=1,\ldots,p,$$
where $d_l= x_l - \bar x$ is the deviation of the $l$-th distortion from the
mean value $\bar x$. Now Eq.(19) can be rewritten as
$$L(\vec\sigma^*,\{x_l\}_1^p)=(q+p\cos w)^2-2\bar x(q+p\cos w)\cos w
-\frac{2(q+p\cos w)}p\sum_{l=1}^p\sigma_ld_l.\eqno(22)$$
Let's discuss the case of slightly different distortions. That is, the
deviations of all the distortions from the mean value $\bar x$ are 
sufficiently small:  
$$\mid d_l\mid <<1,\quad l=1,\ldots,p.$$ 
Then it is clear
that the first two terms give the main contribution in Eq.(22). We would
like to remind that just these two terms were analyzed in \ref{2.1} (see
Eq.(8)). Consequently, 
{\it in the zeroth approximation} we have, that if $\bar x$ is inside the
interval $[x_k,x_{k+1}]$ ($x_k$ are defined by Eqs.(9)), the ground state
belongs to the class $\Sigma_k$, and the global maximum of the functional is
${p\choose k}$-times degenerate. What will happen when the third term in
Eq.(22) is taken into account? It can be expected that for 
sufficiently small $d_l$, this term does not lead the ground state
out of the class $\Sigma_k$, but only lifts of the ground
state degeneracy. Numerical experiments confirm this assumption: when all 
$d_l$ are not very great and $\bar x\in (x_k,x_{k+1})$, {\it all} the local 
maxima of the functional (1) belong to the class $\Sigma_k$, the values of 
the functional at vectors $\vec\sigma^*\in\Sigma_k$ differ slightly and one 
(or several) of these vectors provide(s) the global maximum of the functional.
Thus, we explain one of the possible reasons, why the functional (1) has a lot 
of slightly different local maxima. One frequently faces such situation in the 
neural network theory, Ising model and the optimization problems.

\subsubsection{Other generalizations of the basic model} 
\label{2.2.4}
Without dwelling on the details, we would like to mention other possible 
generalizations of the basic model\footnote{See also Note 3 of \ref{3.2}.}.  
\vskip 2mm
\noindent
${\bf 1^\circ.}$ {\bf The normalized memorized patterns.} 

\noindent
The memorized patterns (4) can be normalized to unit to prevent their length
being dependent on the varying parameter $x$. It is easy to see, that in this
case all the reasonings, which lead to the Theorem, are valid. It can be shown
that here the maximum value of the functional {\it for the ground state}
decreases monotonically as function of $x$. 
\vskip 2mm
\noindent
${\bf 2^\circ.}$ {\bf The identical distortion of a group of coordinates of
the standard.} 

\noindent
Let $m$ coordinates of the standard $\vec \varepsilon (n)$ be distorted 
{\it simultaneously and identically}. 
Then $(p\times n)$-matrix of the memorized patterns has the form
$${\bf S}^{(m)}=
\left(\begin{array}{ccccccccccccc}
1-x&\ldots&1-x&1&\ldots&1&\ldots&1&\ldots&1&1&\ldots&1\\
1&\ldots&1&1-x&\ldots&1-x&\ldots&1&\ldots&1&1&\ldots&1\\
\vdots&\stackrel{m}{\longleftrightarrow}&\vdots&\vdots&
\stackrel{m}{\longleftrightarrow}&\vdots&\cdots&\vdots&
\stackrel{m}{\longleftrightarrow}&\vdots&\vdots&
\stackrel{q}{\longleftrightarrow}&\vdots\\
1&\ldots&1&1&\ldots&1&\ldots&1-x&\ldots&1-x&1&\ldots&1\end{array}
\right).$$
Of course, now $n=p\times m+q$. Let's note that the nondiagonal elements of 
the
relevant connection matrix ${\bf J}^{(m)}$ have the terms, which are quadratic
in distortions.

With regard to the arguments of \ref{2.1}, it is not difficult to show
that the maximum of the functional 
$F^{(m)}=({\bf J}^{(m)}\cdot\vec\sigma,\vec\sigma)$ is achieved only at
"piecewise constant" configuration vectors 
$$\vec\sigma^*=(\underbrace{\sigma_1,\ldots,\sigma_1}_m,\underbrace{
\sigma_2,\ldots,\sigma_2}_m,\ldots,
\underbrace{\sigma_p,\ldots,\sigma_p}_m,\underbrace{1,\ldots,1}_q).$$
The form of these vectors is dictated by the memorized patterns matrix 
${\bf S}^{(m)}$. The value 
of the functional $F^{(m)}$ at the vectors $\vec\sigma^*$ is
$$F^{(m)}(\vec\sigma^*)\propto x^2 - 2x\cos w\left(\frac{q}m+p\cos w\right)+
\left(\frac{q}m+p\cos w\right)^2, $$
where $w$ is the angle between the $p$-dimensional vectors
$\vec\sigma^\prime=(\sigma_1,\sigma_2,\ldots,\sigma_p)$ and 
$\vec\varepsilon (p)$, and $\cos w$ is given by Eq.(7). As above, $2^p$
vectors $\vec\sigma^*$ are distributed among the classes $\Sigma_k^{(m)}$
at which the value of the functional $F^{(m)}$ is constant:
$$F^{(m)}(\vec\sigma^*)=F^{(m)}_k\quad\forall\vec\sigma^*\in\Sigma_k^{(m)},
\quad k=0,1,\ldots,p.$$ 
The structure of the classes $\Sigma_k^{(m)}$ is much alike the structure of
the classes $\Sigma_k$ from \ref{2.1}. The only difference is that here
the value $-1$ have not one of the coordinates $\sigma_l$ of the vector 
$\vec\sigma^*$, but a whole group of $m$ such coordinates. Again, as above, 
the
number of the vectors $\vec\sigma^*$ in the class $\Sigma_k^{(m)}$ is equal to 
$p\choose k$. The same reasoning as in \ref{2.1} leads to the generalization of
the Theorem: 

\noindent
{\it The rebuildings of the ground state from the class 
$\Sigma_{k-1}^{(m)}$
into the class $\Sigma_k^{(m)}$ occur at the points $x_k^{(m)}$ 
$$x_k^{(m)}=\frac{p}{2}\left[1+\frac{q}{q
+mp(\cos w_{k-1}+\cos w_k)}\right]=
p\cdot\frac{\frac{n}m-(2k-1)}{\frac{n}m+p-2(2k-1)},\quad k=1,\ldots,k_{max}.$$
If $p<\frac{q}{2m}+1$, one after another all the $p$ rebuildings of the ground
state take place according to the above scheme: $k_{max}=p$. And if 
$p>\frac{q}{2m}+1$, the last rebuilding is the one whose number is 
$k_{max}=\left[\frac{\frac{n}{m}+p+2}4\right]$. The functional has no other
local maxima.} 

For what follows we would like to note, that when $m$ increases the rebuilding 
points
$x_k^{(m)}$ with $k<\frac{p+1}2$ move to the left, and the rebuilding points 
$x_k^{(m)}$ with $k>\frac{p+1}2$ move to the right. If $p$ is an odd number,
necessarily there is the rebuilding point with the number 
$k=\frac{p+1}2$:\quad $x_{\frac{p+1}2}=p$. When $m$ increases this point
does not move. 

\section{Discussion and Interpretations}
\label{3}
Let us discuss the results which mainly refer to the basic model.

\subsection{Neural networks} 
\label{3.1}
In connection with neural networks the Theorem has to be interpreted in the 
framework of the meaningful setting of the problem, which has
been given in Introduction. Then the Theorem means: {\it The quality of 
"the truth" (i.e., the standard $\vec\varepsilon (n)$) which is understood by
the network depends on the distortion value $x$ during the learning stage and
on the length $p$ of the learning sequence}. We would like to point out some
important details of the suggested interpretation. 

${\bf 1^\circ.}$ In agreement with the common sense the error of the network
increases with the increase of the distortion $x$: when $x$ belongs to the 
interval $(x_k,x_{k+1})$, "the truth" which is understood by the 
network (the class $\Sigma_k$) differs from the standard $\vec\varepsilon (n)$
by $k$ coordinates.

Also it is quite reasonable that the left boundary of the rebuilding region 
$x_1$ is the increasing function of $p$ and $n$. Indeed, when $n$ and $x$ are
fixed, merely due to increase of the number of the memorized patterns $p$ the
value of $x_1$ can be forced to exceed $x$ (of course, if $x$ is not too 
large). As a result $x$ turns out to be on the left of $x_1$, i.e. in the
region where {\it the only} fixed point is the standard $\vec\varepsilon (n)$.
In other words, only 
by an increase of the length $p$ of the learning sequence, one can attain that 
the network understands correctly "the truth" it is tried to be learned.
This conclusion is in agreement with the practical experience according which 
the greater the length of the learning sequence, the better the signal can be 
read through noise. From this point of view the most reliable network of 
the considered type is the one with $p=n$: for any distortions 
$x<x^\prime=\frac{n}2$ it steadily reproduces the standard-vector (see the 
comment to Eq.(10)). 

In the same way when $p$ and $x$ are fixed, merely due to an increase of the 
number $n$ the value of $x_1$ can be forced to exceed $x$. This result is
reasonable too: when $p$ is fixed and the number $n$ increases, the relative
number $p/n$ of the distorted coordinates of the standard-vector decreases.
Naturally, the less the relative weight of the distortion, the better must be
the result of the learning.

${\bf 2^\circ.}$ Under an increase of the distortion parameter not only the 
error of the network increase, but also the number of "noncorrect truths" 
which
are understood by the network: when $x\in(x_k,x_{k+1})$ the number of the
network 
fixed points is $p\choose k$. Such multiple degeneracy of the "truth"
is in agreement with the common idea that in general 
there is only one "real truth", but the number of the "noncorrect truths" is
large. And also, the greater is the distortions at the learning stage, the
more can be the number of "noncorrect truths".

However, when $x$ becomes greater than $x_{\frac{p+1}2}=p$, the number of
"noncorrect truths" decreases. And this number decreases monotonically under
the further 
increase of the distortion parameter. Moreover, according to the Theorem, when
$\frac{p-1}{n-1}<\frac13$, for all $x$ whose values exceed the right boundary
of the rebuilding region, the ground state belongs to the class $\Sigma_p$.
And, in spite of the very great distortion, the ground state is a 
nondegenerate
one! In other words, the number of "noncorrect truths" is a nonmonotone
function of the distortion $x$. 

This result seems to be a paradox. This paradox can be explained when it is
related to the well-known feature of our perception: we interpret 
deviations in the image of a standard {\it as permissible ones only till some 
threshold.} If only this threshold is exceeded, the distorted patterns are 
interpreted as {\it quite different} standard. The Hopfield network under
consideration demonstrate just the same behavior with $x_{\frac{p+1}2}=p$
being the boundary of the permissible distortions of the standard. Let's 
prove the last statement.

Here the notation for the standard $\vec\varepsilon (n)$ differs from the one
used above, and it is
$$\vec\varepsilon^{(+)}=(\underbrace{1,1,\ldots,1}_p,1,\ldots,1).\eqno(23)$$
We also introduce another standard:
$$\vec\varepsilon^{(-)}=(\underbrace{-1,-1,\ldots,-1}_p,1,\ldots,1).\eqno(24)$$
In their not coincident parts the
standards $\vec\varepsilon^{(+)}$ and $\vec\varepsilon^{(-)}$ are opposed with
each other, i.e. they are two opposite statements. Any of the
network fixed points $\vec\sigma^*$ (5) is an intermediate statement between
$\vec\varepsilon^{(+)}$ and $\vec\varepsilon^{(-)}$, which is drawn towards
either one edge of the scale ($\vec\varepsilon^{(+)}$), or the other 
($\vec\varepsilon^{(-)}$). 

Which distortions of the standard $\vec\varepsilon^{(+)}$ have to be assumed 
as
permissible? Evidently, those for which the result of the learning 
resembles the standard $\vec\varepsilon^{(+)}$ more, than its opposition
$\vec\varepsilon^{(-)}$. This is, the fixed points $\vec\sigma^*$  have to
differ from the standard $\vec\varepsilon^{(+)}$ not more than by $\frac{p}2$
first coordinates: $\vec\sigma^*\in \Sigma_k,\mbox{ }k\le\frac{p}2.$ And this
takes place only if $x< x_{\frac{p+1}2}=p$. But if $x>p$, the fixed points 
$\sigma^*$ belong to the class $\Sigma_k$ with $k>\frac{p}2$, and they more 
resemble the standard $\vec\varepsilon^{(-)}$, than $\vec\varepsilon^{(+)}$.

Let's point out the following curious circumstance. Let not the one, but a
whole group $m$ of the coordinates of the standard $\vec\varepsilon^{(+)}$ 
be distorted. Then the rebuilding points $x_k^{(m)}$ with the numbers 
$k<\frac{p+1}2$
move to the left when $m$ increases (see the last paragraph of item $2^\circ$
in \ref{2.2.4}). And this is clear: now the total distortion of the   
standard is $m$-times enlarged and the network earlier ceases to extract the
standard from the memorized patterns. 
It is somewhat unexpected that the boundary $x_{\frac{p+1}2}^{(m)}=p$ of the 
permissible distortions of the standard does not depend on $m$. It might 
appear
that when the total distortion is $m$-times enlarged, the boundary of
the permissible distortions is reached $m$-times quicker. However, this is not
the case. Apparently, the reason is nonadditivity of the influence of the
distortion $x$ on the result of the learning. Out of the simultaneous
distortion of 
$m$ coordinates of the standard, it does not follow at all that $x_k^{(m)}$ 
becomes $m$-times less than $x_k$. When $m$ increases, the rebuilding point 
$x_1^{(m)}$ undergoes the most shift to the left. The shift of $x_2^{(m)}$ is
somewhat smaller, the shift of $x_3^{(m)}$ is still smaller, and so on. The
rebuilding points $x_k^{(m)}$ with the numbers $k \sim p/2$ shift only
negligible. It might be assumed that when the 
distortion $x$ is so big that the result  of the learning differs from the
standard significantly (it is almost impossible to understand the standard),
it is not so important one or several parameters are distorted. Although
formally 
just $x_{\frac{p+1}2}=p$ is the boundary of the permissible distortions.

In our opinion, the most interesting conclusion here is that the Hopfield type
network understands the distortions {\it as permissible} only up to some 
threshold. That is, a nontrivial feature of the human perception is inherent 
in
the artificial network of the examined type.

${\bf 3^\circ.}$ When the distortion $x$ is not very large the memorized
patterns $\vec s^{(l)}$ are interpreted by the network as the distorted
copies of the standard $\vec\varepsilon^{(+)}$. 
But if during the learning stage the distortion $x$ is very large (it exceeds
$p$), the network ceases to understand the memorized patterns (4) as 
distortions
of the standard $\vec\varepsilon^{(+)}$. It does not mean that in this
case the memorized patterns (4) are understood as distortions of the
opposite standard $\vec\varepsilon^{(-)}$: the permissible distortions of 
$\vec\varepsilon^{(-)}$ are the vectors 
$$(\overbrace{-1,\ldots,-1,\underbrace{-1+y}_l,-1,\ldots,-1}^p,1,\ldots,1),
\quad l=1,2,\ldots,p\eqno(25)$$ 
where $y$ is not so large. The question is, what do the memorized patterns (4)
with large distortions $x$ mean? We think, that the increase of $x$ above the
the threshold $p$ can be understood as
the more and more {\it negation} of the standard $\vec\varepsilon^{(+)}$.  As
if the network is learned by the memorized patterns, which {\it denies} the
standard $\vec\varepsilon^{(+)}$. In other words, the network is 
{\it relearned} by presentation of negative examples.

There is big and clear to everybody difference between \underline{the 
relearning} with the help of negative examples and \underline{the learning} of
the opposite truth. In the framework of our model in the last case slightly
distorted $n$-dimensional vectors of the type (25) 
ought to be used as memorized patterns. Then, with no problem the network
understands the standard $\vec\varepsilon^{(-)}$. The relearning is entirely 
other case. It is well known: the better the
incorrect truth has been understood, the more difficult (and sometimes even 
impossible) to correct it; it is comparatively easy to correct the result
slightly, but it is much more difficult to revise it in the main, {\it etc.}

We think, that the dependence of $k_{max}$ on $p$ (see the Theorem) is the
reflection of just these problems. When the number $p$ of the parameters which
have to be corrected is not very great ($\frac{p-1}{n-1}<\frac13$), the
network can be relearned by simple presentation of negative examples. In this
case $k_{max}=p$ and, when "the denial" of the standard 
$\vec\varepsilon^{(+)}$
is rather strong ($x>x_p$), as  "a new" truth the network understands
the opposite standard $\vec\varepsilon^{(-)}$. But if the number of the
corrected parameters is great ($\frac{p-1}{n-1}>\frac13$), to relearn the
network it is not sufficient to present the negative examples. In this case
$\left[\frac{p+1}2\right]<k_{max}<p$
and whatever large $x$ is $(x_{k_{max}}<x<\infty)$, as a new truth the network
understands not the opposite standard $\vec\varepsilon^{(-)}$, but one of the
statements intermediate between $\vec\varepsilon^{(+)}$ and 
$\vec\varepsilon^{(-)}$. Though the understood truth is drawn towards 
$\vec\varepsilon^{(-)}$, since $k_{max}>\frac{p}2$.

Of course, our interpretation is open for discussion. But it seems that in
real life there are a lot of examples, which confirm our 
conception\footnote{In
this connection we are tempted to compare the development of the reforms ("the
relearning") in Russia and in China. In Russia very great number of the
parameters of the social order were simultaneously changed on their 
opposition.
They are a  
very wide group of economic regulations, a group of parameters related to home
and foreign policy, the parameters related to human rights and to the
destruction of  
"the iron curtain" and so on. On the contrary, according to massmedia, in China
only a restricted group of economic regulations has been changed up to now. 

And what are the results? Of course, now Russia is not a semi-fascist Soviet
State, but it does not reach (until now?) the stable State system of the 
modern
democratic type. Already for a long time the situation in Russia is
characterized as a state, which is intermediate between these two oppositions.
The permanent instability and variations of the internal situation in Russia
are 
associated with "the degeneracy of its present ground state". In the same time
(again, if the massmedia information is trustworthy), China demonstrates a
steady increase of well-being of the people. And this stipulates the direction
of its development. 

Being not a specialist, I have no right to intrude into such a special field 
of
knowledge as politology. But in accordance with our results, it appears that, 
may be, "a step by step" strategy of reforms is more prospective. That is, the
result will be better if at first one group of the parameters is corrected
reliably, then the second group, then the third and so on. 
According to the Theorem, taken separately each
group of the parameters can be corrected reliably (of course, if the Theorem
is applicable in this case). On the other hand, the final result of "the
relearning" of these two colossi is unknown yet...}.

\subsection{The Ising model} 
\label{3.2}
The interpretation of the Ising model in 
terms of the matrix $\bf S$ is not known yet. Therefore here the obtained 
results are interpreted starting from the form of the Hamiltonian $\bf J$ 
(3.b). Let's write it in the form    
$$ \bf J\propto\left(\begin{array}{cc}
\bf A&\bf B\\
\bf B^{\rm T}&\bf C\end{array}\right),$$
where the diagonal elements of $(p\times p)$-matrix $\bf A$ and 
$(q\times q)$-matrix $\bf C$ are equal zero, and 
$$\left\{\begin{array}{ll}a_{ij}=1-2y,&i,j=1,2,\ldots,p,\quad i\ne j;\\ 
b_{ik}=1-y,&i=1,2,\ldots,p,\ k=1,2,\ldots,q;\\
c_{kl}=1,&k,l=1,2,\ldots,q,\quad k\ne l;\\
y=\frac{x}p.&\ \end{array}\right.\eqno(26)$$
The matrix $\bf J$ corresponds to a spin system with the infinitely large
interaction radius. The system consists of two subsystems, which are
homogeneous with  
respect to the spin interaction. The interaction between the $p$ spins of the 
first subsystem is equal to $1-2y$; the interaction between the $q$ spins of 
the second subsystem is equal to $1$; the crossinteraction between the spins 
of
the subsystems is equal to $1-y$. When $p=n$ all the spins are interacting 
with each other in the same way and we have the usual mean-field 
approximation for the Ising model at zero temperature.

While $y<\frac12$, all the spins are interacting in the ferromagnetic way; 
when
$\frac12<y$, the interaction between the spins of the first subsystem becomes
of antiferromagnetic type, and when $1<y$ the crossinteraction is of
antiferromagnetic type too. The Theorem allows to trace how the ground state 
depends on the variation of the parameter $y$.

Let $p<n$. For $y\in (-\infty,\frac12)$ the ground state is the ferromagnetic  
one since $\frac12<y_1=\frac{x_1}p$, and for $x<x_1$ the ground state is the
standard-vector $\vec\varepsilon^{(+)}$ (23). It is interesting that the
ground state remains ferromagnetic even if $\frac12<y<y_1$, i.e. when the
antiferromagnetic interactions already are shown up in the system. In other 
words, when $p<n$, there is {\it "a gap"} between the value of $y$ 
corresponding to the destruction of the ferromagnetic  
interaction, and the value of $y$ corresponding to the 
destruction of the ferromagnetic ground state. Only after a "sufficient
amount" of the antiferromagnetic interaction is accumulated, the first
rebuilding  
occurs and the ground state ceases to be the ferromagnetic one. Then,
when another critical "portion" of the antiferromagnetic interaction is 
accumulated, the next rebuilding of the ground state occurs (it happens when
$y$  
exceeds $y_2=\frac{x_2}p$), and so on. After the parameter $y$ reaches the 
value $y^{\prime\prime}=1=\frac{x^{\prime\prime}}p$, the crossinteraction 
becomes antiferromagnetic too. But the rebuildings of the ground state go on,
since usually $x_{k_{max}}> p$ (see the comment to the Theorem in \ref{2.1}). 

The energy $E=-\ F$ {\it of the ground state} as a function of the parameter
$y$ has breaks at the points $y_k=\frac{x_k}p$. It increases till 
$y\le 1$ and decreases when $y>1$. However, if
the memorized patterns are normalized to unit, the energy of the ground state 
is a monotonically increasing function of $y$ (see item $1^\circ$ in 
\ref{2.2.4}).

It is natural to treat $p=const,\mbox{ }n\to\infty$ as an infinitely large
sample with a few number of impurities. In this case all  
$y_k$ stick to the point $y^{\prime\prime}=1$ (see the comment to Eq.(11)).
Depending {\it only on the type of the crossinteraction} between the 
impurities and the rest of the sample, the ground state is either the 
ferromagnetic one, or the spins of the impurities are opposite with respect to
all other spins. 

Finally, let's discuss the case $p=n$. Then all $y_k$ stick to the
point  $y^\prime=\frac12$ (see the comment to Eq.(10)). Here the 
destruction of the ferromagnetic interaction occurs simultaneously with the 
change of the ground state ( "the gap" disappears). As long as the 
interaction of the spins is ferromagnetic ($y<\frac12$), the ground state is 
ferromagnetic too. But when the interaction of the spins becomes 
antiferromagnetic ($y>\frac12$), the ground state turns out to be 
$n\choose \frac{n+1}2$-times degenerate. From the right of $\frac12$ it 
is natural to associate the state of the system with the spin glass phase.

{\bf Note 3.} We would like to remind that the Hamiltonian can be
generalized (see Eqs.(12) and (15)). Nevertheless, at heart of all these
generalizations is the strict relation between the elements of the  
submatrices $\bf A$ and $\bf B$ of the Hamiltonian: 
$$1+a=2b,$$
see Eqs.(26). This restriction on the matrix elements is 
justified by nothing. It would be interesting to examine the case when the
matrix elements $a$ and $b$ do not related in any way. We have succeeded in 
the
generalization of our consideration on this case as well as on the case when 
the linear term $h\sum_{i=1}^n\sigma_i$ has been added to the functional 
$F(\vec\sigma)$ (1). In the Ising model due to such a term the external 
magnetic field can be taken into account. Now we prepare these results for 
publication.

\subsection{Factor analysis} 
\label{3.3}
In the framework of factor analysis a $(p\times n)$-matrix $\bf S$ 
with {\it real} matrix elements (see Eq.(2)) plays an important role.
Here the matrix $\bf S$
is an empirical matrix of the "objects-parameters" which includes the 
exhausting quantitative description of the phenomenon under 
consideration \cite{Brav2}. Its rows are $p$ objects defined by $n$ parameters 
which are available for the measurement:
$$\begin{array}{cll}
\vec s^{(l)}=(s^{(l)}_1,s^{(l)}_2,\ldots,s^{(l)}_n) \in\rm R^n,&
l=1,2,\ldots,p&\mbox{ -- the vector-objects,}\\
\vec s_i=\left(\begin{array}{c}s_i^{(1)}\\s_i^{(2)}\\\vdots\\ s_i^{(p)}
\end{array}\right)\in\rm R^p,&i=1,2,\ldots,n&\mbox{ -- the vector-parameters.}
\end{array}$$
The main idea of the factor analysis is as follows: by proceeding from $n$ 
$p$-dimensional vector-parameters we have to introduce such $t$ artificial 
characteristics, the so called {\it factors}, that their number is {\it much 
less} than $n$ ($t\ll n$), but the description of the objects in terms of 
these
factors does not lead to the loss of an important information about the 
objects. The factors $\vec f_j$ are constructed as linear combinations of 
the vector-parameters $\vec s_i$,
$$\vec f_j=\sum_{i=1}^n a_i^{(j)}\vec s_i,$$
which in the sense of a measure of proximity minimize the total deviation of 
each factor from all the vector-parameters. It has been shown \cite{Lit1}, 
that for the so called {\it centroid method}, which is widespread in 
factor analysis, the problem of the construction of the first factor is 
equivalent to the maximization of the functional (1) with the connection 
matrix
of the Hebb type. The interpretation of the obtained results in terms of  
factor analysis will be presented elsewhere. 

\section{Acknowledgments}
The author is grateful to Prof. Alexander Ezhov for helpful advices on the 
substance of the work and to Dr. Inna Kaganova for preparation of this 
manuscript. Wonderful atmosphere and personal contacts with the members of
workshop "Statistical Physics of Neural Networks" which took place at
Max-Plank-Institute of Physics of Complex Systems (March, 1999, Dresden,
Germany) helped the author in the interpretation of the obtained results.

\section{Appendix}
\subsection{The family of the straight lines $L_k(x)$}
\label{4.1}
The family of the straight lines (8),
$$L_k(x)= (n-2k)^2-2x\frac{p-2k}{p}(n-2k),\quad k=0,1,\ldots,p,$$
is under investigation. When $x=0$ we have $L_k(0)= (n-2k)^2$. Then if  
$p \le n/2$ the values $L_k(0)$ decrease monotonically when $k$ increases:
$$L_0(0)>L_1(0)>L_2(0)>\ldots>L_p(0).$$
And if $p > n/2$, beginning from the number $n-p$ the values $L_k(0)$ 
become twice degenerate:
$$L_0(0)>L_1(0)>\ldots>L_{n-p}(0)=L_p(0)>L_{n-p+1}(0)=L_{p-1}(0)>\ldots $$
Depending on the number $k$ the straight lines $L_k(x)$ 
behave themselves in different ways: for $k<\frac{p}{2}$ and $k>\frac{n}{2}$ 
the slope of $L_k(x)$ is negative, and for $\frac{p}{2}<k<\frac{n}{2}$ it is
positive. The straight lines with the numbers $k=\frac{p}{2}$
and $k^\prime=\frac{n}{2}$ (if such $k$ and $k^\prime$ are found) are parallel
to the axis $OX$.

We denote by $x_{(k,k^\prime)}$ the abscissa of the point of intersection of
the straight lines $L_k(x)$ and $L_{k^\prime}(x)$:
$$x_{(k,k^\prime)}=\frac{p}{2}\left[1+\frac{q}{q+p(\cos w_k+\cos w_
{k^\prime})}\right]=p\frac{n-(k^\prime+k)}{n+p-2(k^\prime+k)},\quad 0\le k<
k^\prime\le p.\eqno(27)$$
We would like to point out that the abscissa 
$x_{(k,k^\prime)}$ depends on the sum $k + k^\prime$ only. When $n+p$ is odd,
the number of {\it different} abscissae 
$x_{(k,k^\prime)}$ is exactly $2p-1$. When $n+p$ is even, there can be such
numbers  
$k^\prime$ and $k$, that $n+p-2(k^\prime+k)=0$. The relevant straight lines
$L_k(x)$ and $L_{k^\prime}(x)$ are parallel (they intersect at the infinity).

Generally, the abscissa $x_{(k,k^\prime)}$ increases when the sum 
$k+k^\prime$ increases
$$x_{(0,1)}<x_{(0,2)}<x_{(0,3)}=x_{(1,2)}<x_{(0,4)}=x_{(1,3)}<\ldots$$
But the increase is strictly monotonic only for
$$\frac{p-1}{n-1}\le\frac13.$$
In this case the denominator of the expression for $x_{(k,k^\prime)}$ is
nonnegative for all $k$ and $k^\prime$, and all the intersections of the
straight lines take place in the region $x>0$. But if
$$\frac{p-1}{n-1} > \frac13,$$
the abscissae $x_{(k,k^\prime)}$ increases monotonically up to the value
$k+k^\prime = \frac{n+p}2$ where it has a discontinuity and becomes negative,
and 
after that $x_{(k,k^\prime)}$ again increases monotonically with increase of 
$k+k^\prime$. 
It is easy to
verify that for $\frac13<\frac{p-1}{n-1}\le\frac12$ all the abscissae
$x_{(k,k^\prime)}$ with $k+k^\prime>\frac{n+p}2$ remain negative. But if
$\frac{p-1}{n-1}>\frac12$, at least $x_{(p-1,p)}$ appears in the region $x>0$.
Nevertheless, in this case also $x_{(p-1,p)}$ remains less than 
$x_{(0,1)}$. In Fig.2 the dependence of $x_{(k,k^\prime)}$ on $k+k^\prime$ is
shown both for $\frac{p-1}{n-1}$ less and greater than $\frac13$.

\begin{figure}[htb]
\begin{center}
\leavevmode
\epsfxsize = 16.2truecm
\epsfysize = 7.5truecm
\epsffile{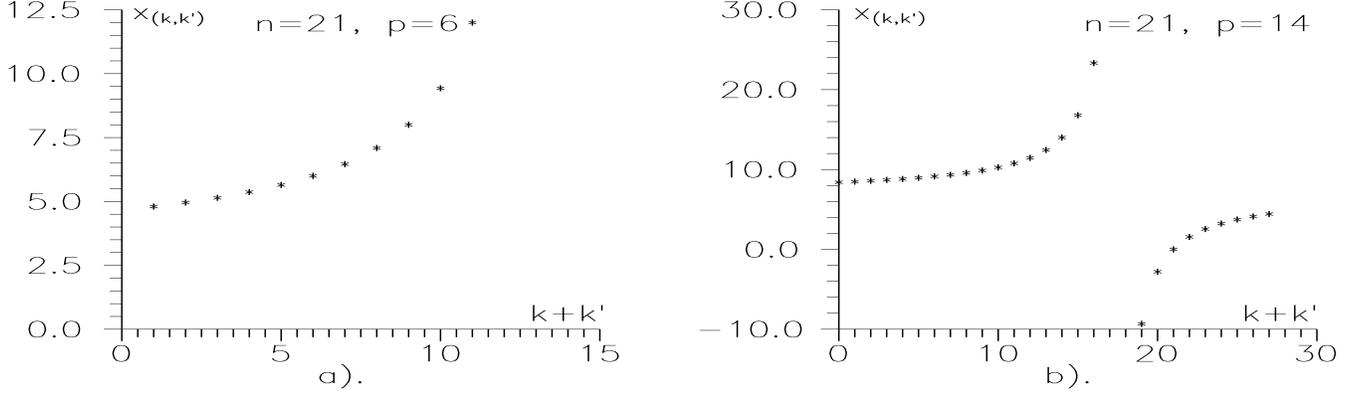}
\caption{The dependence of the abscissa $x_{(k,k^\prime)}$ of the intersection
of the straight lines $L_k(x)$ and
$L_{k^\prime}(x)$ on $k+k^\prime$. Its character is
defined by the value of the parameter $\frac{p-1}{n-1}$:
a). for $\frac{p-1}{n-1}\le\frac13$ the curve $x_{(k,k^\prime)}$ increases
monotonically, remaining all the time in the region $x>0$; 
b). for $\frac13<\frac{p-1}{n-1}$ the monotonically increasing curve 
$x_{(k,k^\prime)}$  
has discontinuity when $k+k^\prime=\frac{n+p}2$, after that again it 
increases monotonically.}
\end{center}
\end{figure}

\subsection{The proof of the Theorem}
We divide the proof into two steps.

${\bf 1^\circ.}$ Let's observe how the ground state changes as $x$ increases.
For $x=0$ the ground state is the one with $k=0$  (see above 
the inequalities for $L_k(0)$), and this state is nondegenerate. As $x$
increases, the state with $k=0$ remains the ground state, till the straight
line $L_0(x)$ does not intersect any other straight line $L_{k^\prime}$. From
Eq.(27) it is
evident, that the first intersection happens with the straight line $L_1(x)$ 
at the point 
$$x_1\equiv x_{(0,1)}= p\frac{n-1}{n+p-2}.$$
After this the ground state is the state with $k=1$ from the class $\Sigma_1$.
This state is $p$-times degenerate.

As $x$ increases further, the state with $k=1$ remains the ground state till
the straight line $L_1(x)$ does not intersect any other straight line
$L_{k^\prime}(x)$. From Eq.(27) it is evident, that the first
intersection takes place with the straight line $L_2(x)$ at the point
$$x_2\equiv x_{(1,2)}= p\frac{n-3}{n+p-6}.$$
After this the ground state is the state with $k=2$ from the class $\Sigma_2$.
And this state is $p\choose 2$-times degenerate.

By analogy with the above reasoning and taking into account the behavior of 
the
family of the straight lines $L_k(x)$ (see \ref{4.1}), we obtain that
when $x$ increases the rebuildings of the ground state happen at the points
$$x_k\equiv x_{(k-1,k)}=\frac{p}{2}\left[1+\frac{q}{q+p(\cos w_{k-1}+
\cos w_k)}\right]= p\frac{n-(2k-1)}{n+p-2(2k-1)}\quad k=1,2,3,\ldots.$$
At the point of the $k$-th rebuilding the $p\choose {k-1}$-times degenerate
ground state from the class $\Sigma_{k-1}$ is replaced by the 
$p\choose k$-times degenerate ground state from the class $\Sigma_k$.

The rebuildings of the ground state take place till the increase of $x$ is
accompanied by the intersections of the straight lines $L_{k-1}(x)$ and 
$L_k(x)$. In other words, till $2k-1$ is less than $\frac{n+p}2$. Since for
$\frac{p-1}{n-1}\le\frac13$ even $2p-1 < \frac{n+p}2$, in this case as $x$
increases one after the other all the $p$ possible rebuildings of the ground
state take place. Consequently, for all $x > x_p \equiv x_{(p-1,p)}$, the
ground state remains in the class 
$\Sigma_p$. (Strictly speaking, if $\frac{p-1}{n-1}=\frac13$, the last
rebuilding of the ground state happens at the infinity.)

But if $\frac{p-1}{n-1}>\frac13$, the rebuildings of the ground state come to
an end when $2k-1$ becomes more than $\frac{n+p}2$. For these $k$ the abscissa
$x_{(k-1,k)}$ of the intersection of the straight lines $L_{k-1}$ and $L_k$ 
is negative, and under the further increase of $x$ there is no new 
intersection
of the straight line $L_{k-1}$ with any other straight line $L_{k^\prime}$. In
other words, the state with the number $k-1$ remains the ground state under 
the
unlimited increase of $x$. In this case the number $k_{max}$ of the last
rebuilding is defined from the equation
$$2k_{max}-1=\frac{n+p}2\Longrightarrow k_{max}=\left[\frac{n+p+2}4\right].
\eqno(28)$$
Thus, in the region $x > 0$ the last rebuilding of the ground state takes 
place
when $x=x_{k_{max}}$. (If $\frac{n+p+2}4=\left[\frac{n+p+2}4\right]$, the
last rebuilding takes place at the infinity.) For negative $x$ there are no 
rebuildings of the ground state, since in this region the straight line
$L_0(x)$ majorizes all other straight lines.

${\bf 2^\circ.}$ To complete the proof, let's show that the functional has no
other local maxima. The configuration vector, which provides a local maximum,
has the form (5) necessarily. Let for some $x$ this vector belong to the class
$\Sigma_k$. Then the inequalities 
$$L_k(x)\ge L_{k-1}(x),L_{k+1}(x)$$
must be fulfilled. Removing the brackets and collecting the similar terms we
obtain:  
$$x[2\pm(n+p-4k)]\ge p[1\pm(n-2k)].\eqno(29)$$

A). Let $n+p-4k\ge 2$. It means that $k+1\le\frac{n+p+2}4$, or, in other 
words,
that $k\in [1,k_{max}-1]$. The sign "+" in Eq.(29) results in the inequality 
$$x\ge p\frac{n-(2k-1)}{n+p-2(2k-1)}=x_k,$$
and the sign "--" results in the inequality
$$x\le p\frac{n-(2k+1)}{n+p-2(2k+1)}=x_{k+1}.$$
Thus, for $k\in [1,k_{max}-1]$ the state with the number $k$ is the local
maximum only if $x$ belongs to the interval $[x_k,x_{k+1}]$, where this state
is the global maximum.

B). Let now $-2\le n+p-4k<2$. It means that $k>k_{max}-1$ and $k\le k_{max}$;
in other words, that $k=k_{max}$. 
The sign "+" in Eq.(29) results in the inequality $x\ge x_{k_{max}}$, and the
sign "--" results in the inequality $x\ge x_{k_{max}+1}$. But we know that 
the
value of $x_{k_{max}+1}$ is already negative (see above \ref{4.1}).
Consequently, the simultaneous fulfillment of the last two inequalities takes 
place in the region $x\ge x_{k_{max}}$. And this means that the state with 
the number $k_{max}$ is the local maximum of the functional only if it is its 
global maximum.

C). And, finally, let $n+p-4k<-2$; in other words, let $k>k_{max}$.
The sign "+" in Eq.(29) results in the inequality $x\le x_k$, and the sign
"--" results in the inequality $x\ge x_{k+1}$. But as it has been shown in 
\ref{4.1}, even for $k>k_{max}$ we have the ordering $x_k < x_{k+1}$. 
Consequently, for $k>k_{max}$ the simultaneous fulfillment of 
inequalities (29)
is impossible. This completes the proof.

\subsection{The region of the ground state rebuilding}
The region where the parameter $x$ changes from $x_1$ to $x_{k_{max}}$, will
be called {\it the rebuilding region} of the ground state. Let's examine this
region. 

The left boundary of the rebuilding region is 
$$x_1=p\cdot\frac{n-1}{n+p-2}\ge\frac{p}2,$$
where the equality takes place only for $p=n$. It is easy to verify that 
$x_1$
is a monotone increasing function of both $p$ and $n$. These statements are
important for the meaningful interpretation of the obtained results.

The expression for the right boundary of the rebuilding region depends on the
relation between $p$ and $n$. If $\frac{p-1}{n-1}<\frac13$, then $k_{max}=p$
and 
$$x_{k_{max}}=x_p=p\cdot\frac{n+1-2p}{n+2-3p}.$$
As function of $n$ the value of $x_p$  decreases. Consequently, for 
$n\to\infty$ and fixed $p$ the rebuilding region contracts to the point
$x^{\prime\prime}=p$. And if $\frac{p-1}{n-1}>\frac13$, than 
$$x_{k_{max}}=p\cdot\frac{n-p+4\kappa}{8\kappa},$$
where $4\kappa=1,2\mbox{ or }3$, and $k_{max}$ is given by Eq.(28). 
When $n$ is fixed and $p\to n$, the rebuilding region also contracts to the
point. It is not difficult to see that this point is $x^\prime=\frac{n}{2}$.

In many respects the point $x^{\prime\prime}=p$ is remarkable (see \ref{3}).
It is easy to see that before $x$ becomes equal to
$x^{\prime\prime}=p$, exactly $\frac{p+1}2$ rebuildings of the ground state
take place. This result does not depend on $n$ (if only $p$ is less than
$n-1$). 

\subsection{The rotation of the memorized patterns and $k_{max}$}

Let's prove the generalization of the Theorem for the case, when all the
memorized patterns (4) are subjected to the rotation as a whole (see
\ref{2.2.2}). For simplicity we assume that the number $t+1$ of different
values of the cosines (17) is even: 
$$t+1=2r.$$ 
Let's agree the subscript $k$ to take the values $1, 2,\ldots, r=\frac{t+1}2$ 
only.

Since for every $k$ we have $\cos w_k=-\cos w_{t+1-k}$, then 
$\cos w_{r-1}=-\cos w_r$ and the ordered sequence of the cosines has the form:
$$\cos w_0>\cos w_1>\ldots>\cos w_{r-1}>0>\cos w_r>\cos w_{r+1}>
\ldots>\cos w_{2r-1}.$$
It is not difficult to prove that the rebuilding of the ground state from the
class $\Sigma_{j-1}^{(U)}$ into the class $\Sigma_j^{(U)}$ happens at the point
$x_j$ (18). In what follows we'll prove that 
{\it if $x_1>\frac34p$, one after
another all the $t$ rebuildings  
of the ground state take place according to the above scheme: $k_{max}=t$. And
if $x_1<\frac34p$, the rebuildings of the ground state come to an end when the 
denominator in Eq.(18) becomes negative.}

The numbers $k$ and $t+1-k$ are symmetric with regard to the ends of the
sequence of the rebuilding points. For every rebuilding point 
$$x_k=\frac{p}2\cdot \left[1+\frac{q}{q+p(\cos w_{k-1}+\cos w_k)}\right],$$  
the point $x_{t+1-k}$ is called {\it the dual} rebuilding point. The 
expression for $x_{t+1-k}$,
$$x_{t+1-k}=\frac{p}2\cdot \left(1+\frac{q}{q-p(\cos w_{k-1}+
\cos w_k)}\right),$$
differs from the expression for $x_k$ by the sign in the denominator only.
The rebuilding point 
$$x_r=\frac{p}2\cdot \left[1+\frac{q}{q+p(\cos w_{r-1}+\cos w_r)}\right]=
\frac{p}2\cdot \left[1+\frac{q}q\right]=p$$
is dual to itself. We begin the proof of the general statement with the proof
of the following 

\noindent
{\bf Lemma.}

\noindent
{\it 1. The inequality $x_k>\frac{3}4p$ is equivalent to the inequality 
$q-p(\cos w_{k-1}+\cos w_k)>0.$}

\noindent
{\it 2. For the value of $x_{t+1-k}$ to be negative, it is necessary and 
sufficient to have $\frac{2}3p<x_k<\frac{3}4p.$}

The proof of the first statement is obtained from the convertible sequence of
the inequalities 
$$x_k>\frac{3}4p\Longleftrightarrow\frac{q}{q+p(\cos w_{k-1}+\cos w_k)}>
\frac{1}2\Longleftrightarrow q-p(\cos w_{k-1}+\cos w_k)>0.$$ 
Let's prove the second statement. Let 
$$x_{t+1-k}=\frac{p}2\cdot \left[1+\frac{q}{q-p(\cos w_{k-1}+
\cos w_k)}\right]<0.$$
Then, firstly, $q-p(\cos w_{k-1}+\cos w_k)<0$, and it is equivalent to the
inequality $x_k<\frac{3}4p$. And, secondly:
$$\frac{q}{q-p(\cos w_{k-1}+\cos w_k)}< -1,$$
from where we directly obtain $q+p(\cos w_{k-1}+\cos w_k)<3q$, and, 
consequently, 
$$x_k=\frac{p}2\cdot \left[1+\frac{q}{q+p(\cos w_{k-1}+\cos w_k)}\right]>
\frac{p}2\cdot \left[1+\frac{q}{3q}\right]=\frac{2}3p.$$
Assume the converse. Let $\frac{2}3p<x_k<\frac{3}4p$. This means that 
$$-q<q+p(\cos w_{t-k}+\cos w_{t-k+1})<0,$$
from where it follows that
$$x_{t+1-k}=\frac{p}2\cdot \left[1-\frac{q}{\mid q+p(\cos w_{t-k}+
\cos w_{t-k+1})\mid}\right]<\frac{p}2\cdot \left[1-\frac{q}q\right]=0,$$
i.e. that the abscissa $x_{t+1-k}$ is negative. This concludes the proof of 
the Lemma.

Generally, the first rebuilding point $x_1$ always exceeds $\frac{p}2$.
The proof of the main statement is in three steps.

\noindent
{\bf Step 1.} Let's examine the case $x_1>\frac{3}4p$.
Then according to the first conclusion of the Lemma 
$q-p(\cos w_0+\cos w_1)>0$ and therefore a similar inequality is true for any
other $k$: $q-p(\cos w_{k-1}+\cos w_k)>0$. Consequently, all the rebuilding
points $x_j$ (18) are positive. And this is just the case 
when in consecutive order one after another all $t$ possible
rebuildings of the ground state take place. It is interesting that this
case corresponds to very simple requirement to the first rebuilding point:
$x_1>\frac{3}4p$.   

\noindent
{\bf Step 2.} Now we investigate the case $\frac{2}3p<x_1<\frac{3}4p$. 
According to
the second statement of the Lemma, here the rebuilding point $x_t$, 
which is dual to $x_1$, is negative: $x_t<0$. Then we pass from $x_1$ to $x_2$:
$x_1 < x_2$. If $x_2$ is still less than $\frac{3}4p$, its dual rebuilding
point $x_{t-1}$ is negative too. Then we must to pass from $x_2$ to
$x_3$: 
if $x_3$ is also less than $\frac{3}4p$, its dual
rebuilding point $x_{t-2}$ is also negative, and so on. Since at least
$x_r=p>\frac{3}4p$, 
sooner or later we get the number $k$, for which $x_k$ exceeds 
$\frac{3}4p$: $x_k>\frac{3}4p$. Then according to the Lemma its dual rebuilding
point $x_{t+1-k}$ is already positive:
$$x_{t+1-k}=\frac{p}2\cdot 
\left[1+\frac{q}{q-p(\cos w_{k-1}+\cos w_k)}\right]>0.$$ 
It is evident that $x_{t+1-k}$ is not only positive, but exceeds $x_1$:
$$x_{t+1-k}-x_1\sim \cos w_0+\cos w_1+\cos w_{k-1}+\cos w_k>0.$$
Thus, the last rebuilding of the ground state happens at the point with
the number $t+1-k$. The next rebuilding point, $x_{t-k+2}$, is already
negative. In other words, $k_{max}$ is exactly equal to $j\equiv t+1-k$, where 
$$q+p(\cos w_{j-1}+\cos w_j)>0,\mbox{ but }q+p(\cos w_j+\cos w_{j+1})<0.$$
And just these inequalities give the statement we have to prove.
 
\noindent
{\bf Step 3.} Finally, we have to examine the case $x_1<\frac{2}3p$. This 
inequality
is equivalent to $q-p(\cos w_0+\cos w_1)<-q$, from which it
follows that $x_t>0$:
$$x_t=\frac{p}2\cdot \left[1-\frac{q}{\mid q+p(\cos w_0+\cos w_1)
\mid}\right]>\frac{p}2\cdot \left[1-\frac{q}q\right]=0.$$
In the same time $x_t<x_1$:
$$x_1-x_t\sim
\frac{q}{q+p(\cos w_0+\cos w_1)}-\frac{q}{q-p(\cos w_0+\cos w_1)}>0.$$ 
We pass from $x_1$ to $x_2$. If $x_2$ is also less than $\frac{2}3p$, its dual
rebuilding point $x_{t-1}$ is also positive (but this time  
it does not exceed $x_t$). Increasing the number $k$ we sooner or later get
the rebuilding point $x_k$ which exceeds $\frac{2}3p$. And then we
find ourselves under the conditions discussed in Step 2: 
$\frac{2}3p<x_k<\frac{3}4p$. Here
the point $x_{t+1-k}$, which is dual to $x_k$, is negative. At the next steps
we get such a number $k$, that  
$x_k$ exceeds $\frac{3}4p$. According to the above reasoning this means that
$k_{max}=t+1-k$. In other words, in the case $x_1<\frac{2}3p$, at first 
$x_j$ increase up to the maximal positive value, then become negative, and
after that they increase. Beginning with a certain 
number $j$, the values of $x_j$ are positive again, but less than $x_1$. This
concludes the proof.

Thus, the rebuilding points $x_j,\mbox{ }j=1,2,\ldots,t$ behave themselves in 
different ways depending on the interval to which the first point $x_1$
belongs: 
$$x_1>\frac{3}4p,\quad\frac{2}3p<x_1<\frac{3}4p,\quad x_1<\frac{2}3p.
\eqno(30)$$
The analogous result has been also obtained above, when in \ref{4.1}
the family of the straight lines $L_k(x)$ was investigated. There the behavior
of the rebuilding points were depended on the belonging of the ratio
$\frac{p-1}{n-1}$ to one of the three intervals 
$$\frac{p-1}{n-1}<\frac{1}3,\quad\frac{1}3<\frac{p-1}{n-1}<\frac{1}2,
\quad\frac{1}2<\frac{p-1}{n-1}.\eqno(31)$$
It is easy to verify that for $u_1=u_2=\ldots=u_p=1$ the inequalities (30)
turn into inequalities (31). The case of an even $t=2r$ is examined in the 
same way.

\end{document}